\documentclass{elsart}
\usepackage{graphicx}
\usepackage{amssymb}
\begin{document}

\begin{frontmatter}

% Title, authors and addresses

% use the thanksref command within \title, \author or \address for footnotes;
% use the corauthref command within \author for corresponding author footnotes;
% use the ead command for the email address,
% and the form \ead[url] for the home page:
% \title{Title\thanksref{label1}}
% \thanks[label1]{}
%\author{Name\corauthref{cor1}\thanksref{label2}}
%\ead{joshua.spitz@yale.edu}
% \ead[url]{home page}
% \thanks[label2]{}

%\address{Address\thanksref{label3}}
% \thanks[label3]{}

\title{A Regenerable Filter for Liquid Argon Purification}
\
% use optional labels to link authors explicitly to addresses:
% \author[label1,label2]{}
% \address[label1]{}
% \address[label2]{}

\author[Yale]{A. Curioni},
\author[Yale]{B.T. Fleming},
\author[FNAL]{W. Jaskierny},
\author[FNAL]{C. Kendziora},
\author[FNAL]{J. Krider},
\author[FNAL]{S. Pordes},
\author[Yale]{M. Soderberg},
\author[Yale]{J. Spitz\corauthref{cor1}},
\ead{joshua.spitz@yale.edu}
\author[FNAL]{T. Tope},
\author[Yale]{T. Wongjirad}
\address[FNAL] {Particle Physics Division, Fermi National Accelerator Laboratory, Chicago, Illinois, USA}
\address[Yale]{Department of Physics, Yale University, New Haven, Connecticut, USA}
\corauth[cor1]{Corresponding author.}
\begin{abstract}
A filter system for removing electronegative impurities from liquid argon is described.  The active components of the filter are adsorbing molecular sieve and activated-copper-coated alumina granules.  The system is capable of purifying liquid argon to an oxygen-equivalent impurity concentration of better than 30~parts~per~trillion, corresponding to an electron drift lifetime of at least 10~ms.  Reduction reactions that occur at $\sim$250$^\circ$C allow the filter material to be regenerated in-situ through a simple procedure.  In the following work we describe the filter design, performance, and regeneration process.    
\end{abstract}

\end{frontmatter}

\section{Introduction}
\label{sec:intro}
Liquid Argon Time Projection Chambers (LArTPCs)~\cite{rubbia} feature bubble-chamber-like-quality three dimensional tracking and total absorption calorimetry for high efficiency and low background neutrino detection. Sensitivity to $\theta_{13}$ and leptonic CP-violation with an accelerator-based long baseline neutrino oscillation experiment will require a multi-kiloton scale far detector~\cite{modular}\cite{landd}\cite{glacier}\cite{flare}. While LArTPCs of up to 600~tons~\cite{Icarus:2004} have been successfully built and operated, further R\&D is necessary for a multi-kiloton detector. 

One of the most significant challenges facing large LArTPCs is achieving and maintaining a high level of liquid argon purity. Ionization electrons, created by neutrino-interaction-induced tracks, are drifted along electric field lines to readout electrodes in a LArTPC. Even in a modularized design, an electron drift distance on the order of 5~m is necessary for a realistic multi-kiloton scale detector. Drifting electrons over such lengths requires a many millisecond lifetime given a drift velocity of 1.5~mm/$\mu$s with a typical applied field of 500~V/cm~\cite{walk}. For these lifetimes, the concentration of electronegative contaminants in the liquid must be kept to the 10's of ppt level to prevent excessive signal degradation through electron attachment. In the following, we describe a method for electronegative impurity removal through a non-proprietary filter where previous experiments have used primarily proprietary filters. The advantage of this system is that the filters can be regenerated and reused in-situ, an important feature for multi-kiloton scale experiments. 
\begin{figure*}[htb]
\centering
\includegraphics[width=5.6in]{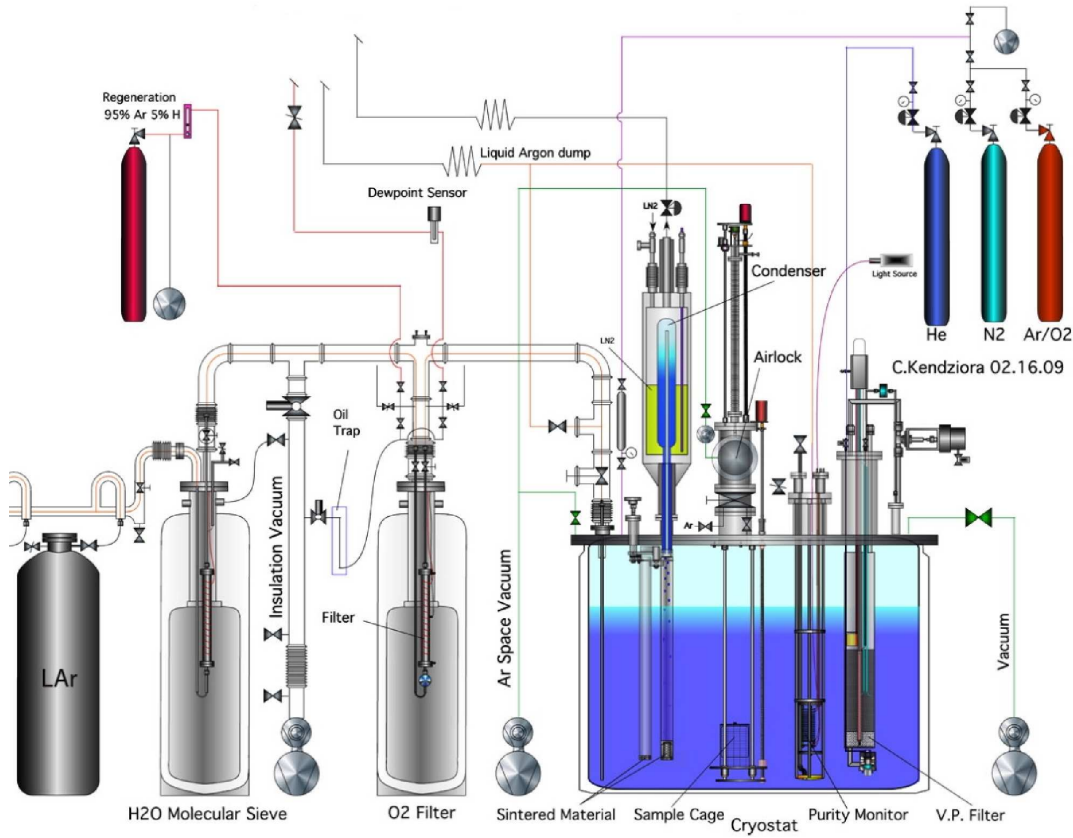}
\caption{The Materials Test System at FNAL. Liquid argon used to fill the cryostat flows from left to right in the schematic.}
\label{fig:Luke}
\end{figure*}
\label{section:FNALfilter}
\section{Filter System Design and Operation}
\subsection{MTS Filter Design and Implementation}
The argon filter concept, centered around impurity removal in the liquid phase, is based on work done by the ICARUS collaboration~\cite{Benetti1993}. This technique was first utilized  with the ICARUS 3~ton LArTPC~\cite{3ton1}\cite{3ton2} and for the first time in an accelerator-based neutrino beam with the ICARUS 50~L LArTPC~\cite{larneutrino}. 

The filters described here are comprised of activated-copper-coated granules~\cite{engelhard} and adsorbent molecular sieve~\cite{sigma}, each of which can be regenerated in-situ. They have been built at Fermi National Accelerator Laboratory (FNAL) and were used by groups at FNAL and Yale University in liquid argon test stands at each location. 

Two different filter systems have been constructed at FNAL for use in the facility's Materials Test System (MTS).  The main purposes of the MTS are to provide a system in which the effect of different materials on liquid argon purity can be observed and to test filtration techniques. A schematic of the MTS is given in Figure~\ref{fig:Luke}.  

Compared to the 10's of parts~per~trillion (ppt) needed for large LArTPCs, commercial-grade liquid argon has an oxygen-equivalent impurity concentration of approximately 1~part~per~million. The first of the two systems installed in the MTS consists of an H$_2$O-removing molecular sieve filter and an activated-copper O$_2$-removing filter; this system is used to purify commercial-grade liquid argon as it flows into the 250~L cryostat. 

\begin{figure}[htb]
\centering
\includegraphics[width=3.0in]{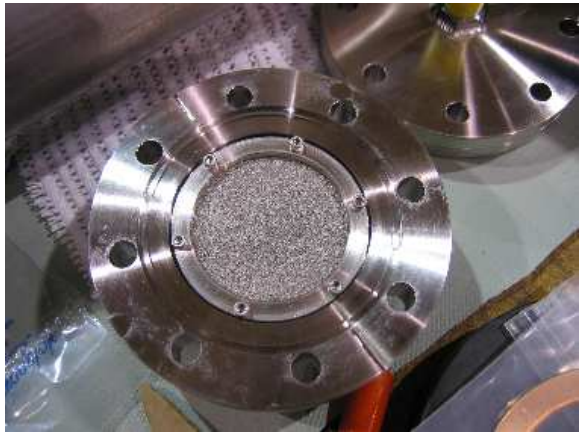}
\caption{The filter's end cap flange, sintered metal disk and holding ring assembled.}
\label{fig:bottomassembly}
\end{figure}

\begin{figure}[htb]
\centering
\includegraphics[width=3.0in]{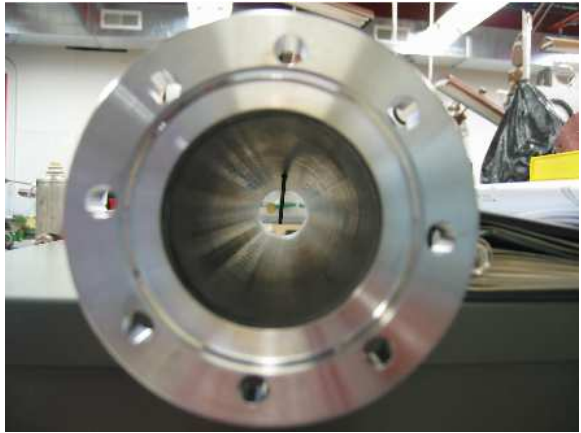}
\caption{View down the inside of an empty filter with thermocouple visible.}
\label{fig:thermocouple}
\end{figure}

The two filters comprising the first system are constructed identically and both sit inside evacuated vessels. The canister holding the filter material is made with a 24'' long stainless steel 2-3/8'' diameter tube capped at both ends by 4-1/2'' outer diameter Conflat flanges. Input and exhaust ports connected to the Conflat flanges are made with  3/8'' stainless steel tubing. Sintered metal disks are clamped to both filter ends. An input-side disk provides flow resistance, allowing liquid to spread throughout the entire filter volume. Both disks prevent filter dust from leaving and keep the granules contained.  A disk, clamping ring, and flange are shown assembled in Figure~\ref{fig:bottomassembly}. Stainless steel tubes and valves are attached to input and output lines to allow evacuation and/or purging with argon gas. A thermocouple is installed inside to monitor filter temperature during the regeneration process.  Figure~\ref{fig:thermocouple}  shows a filter canister with thermocouple visible. Heating tape, used to increase filter temperature during regeneration, is wrapped around the main body and clamped in place by stainless steel tubing. Regeneration is performed ``in-situ'', meaning that the filters are regenerated where they sit in the supply system. Building this into the filter design makes the filter more complicated and the operation of the regeneration and the supply system more simple.

\begin{figure}[htb]
\centering
\includegraphics*[width=3.2in,viewport=120 60 630 550]{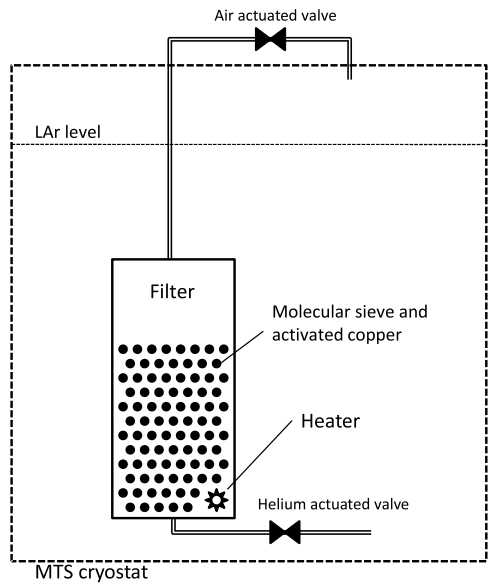}
\caption{The FNAL vapor pump filter inside the MTS cryostat.}
\label{fig:FNALfilter}
\end{figure}

A second filter system, called the ``vapor pump'' filter, is installed directly inside the MTS cryostat.  The purpose of this filter is to remove impurities from test materials introduced inside the cryostat.  The vapor pump filter is constructed out of a 33'' long stainless steel tube with an outer diameter of 5-3/4''. A port at the bottom of the filter is controlled by a normally open He-actuated bellows valve.  Another port, controlled by an air-actuated bellows valve, is installed at the top of the filter and connects the filter volume to the space above the cryostat's liquid level. The main body of the filter contains about 3.0~kg of activated-copper-coated alumina granules and 0.7~kg of molecular sieve material. Heaters and thermocouples are located in the center of the filter in order to heat the granules and monitor the temperature, respectively. For regeneration, the entire vapor pump filter assembly is removed from the cryostat through the aperture of an 8'' Conflat flange. A schematic of the FNAL vapor pump filter is given in Figure~\ref{fig:FNALfilter}.

It is worth noting that we have not performed a rigorous filter geometry study. However, we find that the filters with the dimensions above are reasonable for use in the MTS with consideration for fill rate, the liquid argon supply's inherent impurity concentration, and the total volume of liquid required.

\vspace{1cm}

A filter-cryostat volume exchange proceeds as follows: Liquid is allowed into the filter by opening the top and bottom ports to the cryostat. The top valve is then closed, and the heater inside the filter is activated.  Since the filter volume is closed off at the top, gas created inside the filter cannot escape.  The pressure builds up and forces liquid out of the bottom port. Once all the liquid has been expelled, the top valve is opened and the trapped argon gas inside is released into the cryostat.  The pressure inside the filter equilibrates with the gas pressure at the top of the cryostat and liquid outside the filter flows in. A complete cycle takes approximately two minutes, one minute for emptying and one minute for filling. The vapor pump filters one cryostat volume in approximately one hour, depending on the cryostat liquid level.  

\subsection{Filter Regeneration}
\begin{figure*}[htb]
\centering
\includegraphics[width=4.7in,viewport=55 5 450 260]{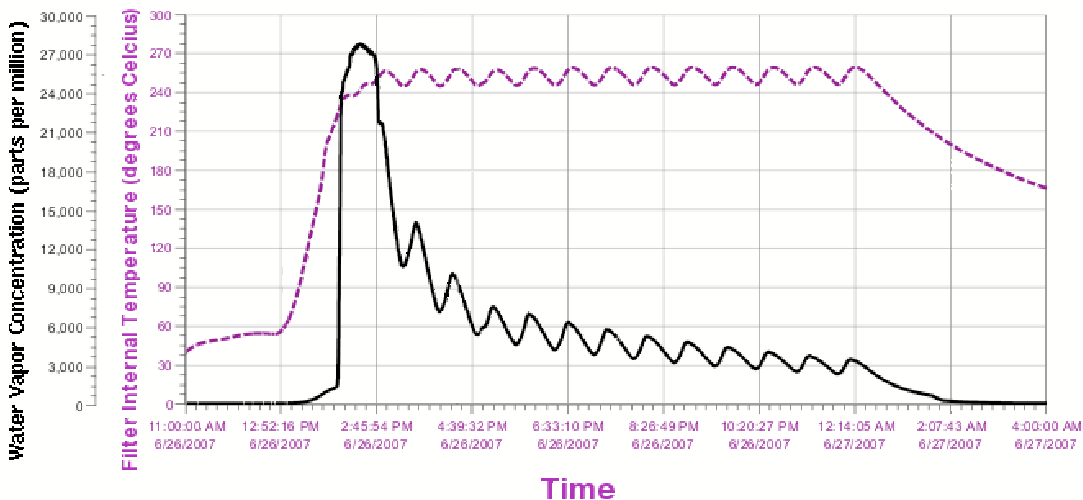}
\vspace{-.6cm}
\caption{Filter temperature (dashed line) and water vapor concentration (solid line) during regeneration.}
\label{fig:regen}
\end{figure*}

Accumulated electronegative impurities reduce the effectiveness of the filter material. It is our practice to regenerate a filter after the passage of $\sim$1000~L of liquid argon. The activated-copper-containing filters are regenerated in accordance with the manufacturer's instructions~\cite{engelhard}. The filters are heated to 250$^\circ$C while a 95:5 mixture of Ar:H$_2$ gas flows through. The gas removes impurities through reduction reactions.
 A dew point monitor, installed at the exhaust end of a filter, is used to monitor the regeneration process.  Figure~\ref{fig:regen} shows the water vapor concentration and temperature inside a filter over the course of a regeneration run.  
The water vapor concentration follows a characteristic rise (above ambient concentration), due to the initial reduction reactions on the granules, and asymptotic fall, as the impurity concentration output gets lower and lower. After about 900 volume changes of gas through a filter (the maximum flow rate of 5~ft$^3$/hour is set by the gas delivery system) the gas flow is turned off. A vacuum pump is then used to remove additional water vapor for 24~hours.  Regeneration of the pure molecular-sieve filter is accomplished with baking 10~hours to 250$^\circ$C while evacuating with a vacuum pump. After the regeneration procedure, the filters are maintained with positive pressure argon gas.
\vspace{-.5cm}
\subsection{MTS Filling Procedure}
The MTS is filled according to the following procedure: A commercial-grade dewar of liquid argon is connected to the input line of the MTS. The argon is allowed to flow through the system, after which it is diverted out of a valve just before the input to the cryostat. When liquid appears at this output, the cryostat is isolated from its vacuum pump, the diversion outlet closed and argon allowed to enter the cryostat. This procedure is adopted with the idea that any residual contaminants in the filter system are removed by the initial argon flow. The cryostat is filled at a rate of 1.0~LPM while venting to atmosphere.   

\subsection{Measuring the Concentration of Electronegative Impurities}

Filter performance is determined using a liquid argon purity measurement inside of the cryostat.  To measure purity, we use the ICARUS-style purity monitor described in~\cite{carugno}. Ultraviolet light from a xenon flash lamp is guided into the purity monitor through a solarization-resistant quartz optical fiber~\cite{polymicro}.  The light impinges upon a photocathode (Au at FNAL, GaAs at Yale), producing photoelectrons at the cathode of the purity monitor.  An electric field is used to drift the electrons along the length of the purity monitor to an anode.  The electron drift lifetime is found by measuring the amount of charge that reaches the anode, the initial charge that has left the cathode, and the electron drift time. The collected charge at the cathode and anode are displayed on a digital oscilloscope in order to measure the pulse heights.  The electron drift lifetime, $\tau_{lifetime}$, is found with
\begin{equation}
\label{eq:lifetime}
\frac{Q_{anode}}{Q_{cathode}} = e^{-t/\tau_{lifetime}}
\end{equation}
where $Q_{anode}$ and $Q_{cathode}$ are taken to be proportional to the pulse heights measured by the electronics readout and $t$ is the drift interval. Figure~\ref{fig:ScopeWaveform} is a typical oscilloscope display of pulses from the purity monitor's cathode and anode electronics readout at the MTS.  In this particular event, the cathode and anode signals are separated by $t=\mathrm{5.7}$~ms ($\tau_{lifetime}=\mathrm{10}$~ms).
\section{Results and Discussion}
\subsection{Electron drift lifetime and impurity concentration}
\begin{figure}
\centering
\includegraphics[width=3.0in]{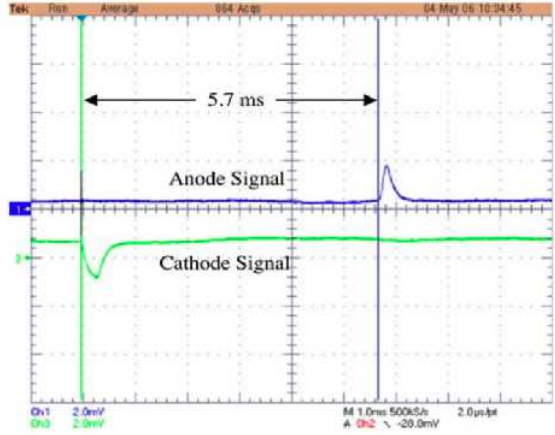}
\caption{Typical signal waveforms from the purity monitor's cathode and anode.  The pulses are separated by a 5.7~ms drift time.}
\label{fig:ScopeWaveform}
\end{figure}

Electron drift lifetimes of $\sim$10~ms, as displayed in Figure~\ref{fig:ScopeWaveform}, are routinely observed by the purity monitor in the MTS.  Note that these values are at the upper limit of what the instrument can measure, as the MTS purity monitor is only 17~cm in length.  A longer purity monitor is being tested in order to measure longer lifetimes.

From~\cite{Bakale1976}, an estimate of the electronegative impurity concentration can be found with 
\begin{equation}
\label{eq:electronDecay}
d[e]/dt = -k_{s}[S][e]
\end{equation}
where $[e]$ is the electron concentration, $[S]$ is the electronegative impurity concentration, and $k_{s}$ is the electron attachment rate constant. Electron attachment due to neutralization at the anode and recombination with positive ions are ignored. At electric fields of 15-100~V/cm -- the fields with which the lifetime measurements are made -- the rate constant of attachment, $k_{s}$ ($\approx$10$^{11}$~M$^{-1}$s$^{-1}$~\cite{Bakale1976}), is weakly dependent on electric field strength~\cite{IcarusPurity:2004}.  

A solution to Equation~\ref{eq:electronDecay} is
\begin{equation}
\label{eq:e_solution_concen}
[e(t)] = [e_{0}]e^{-k_{s}[S]t}
\end{equation}
which is equivalent to 
\begin{equation}
\label{eq:e_solution}
\frac{Q_{anode}}{Q_{cathode}} = e^{-k_{s}[S]t}
\end{equation}
where $Q_{cathode}$ and $Q_{anode}$ are related to $[e(t)]$ and $[e_{0}]$, respectively, by volume and unit conversion factors.  Comparison between Equations \ref{eq:lifetime} and \ref{eq:e_solution} allows one to relate the electron drift lifetime to the electronegative impurity concentration by 
\begin{equation}
\label{eq:lifetimeToM}
[S] = (35.8 \times k_{s} \times \tau_{lifetime})^{-1}
\end{equation}
where the molar volume of liquid argon factor of 35.8 is needed to convert the units of $[S]$ (moles~per~liter) to a molar fraction given in ppt.  As we are using the rate constant for oxygen, impurity level is given in an oxygen-equivalent concentration.  With an electron drift lifetime of 10~ms in liquid argon and $k_{s}=\mathrm{10}^{11}$~M$^{-1}$s$^{-1}$, the corresponding oxygen-equivalent concentration of electronegative impurities is about 30 ppt. 

\subsection{LArTPC Tracks}
\begin{figure}
\centering
\includegraphics[width=2.7in]{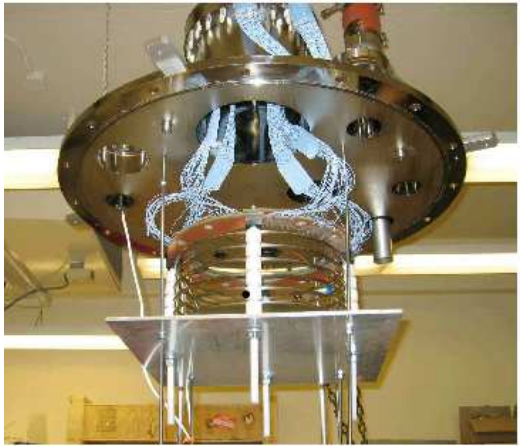}
\caption{The Yale LArTPC.  The TPC is shown suspended from the top flange.}
\label{fig:YaleTPC}
\end{figure}

\begin{figure}
\centering
\includegraphics[width=1.2in,viewport=260 40 520 630]{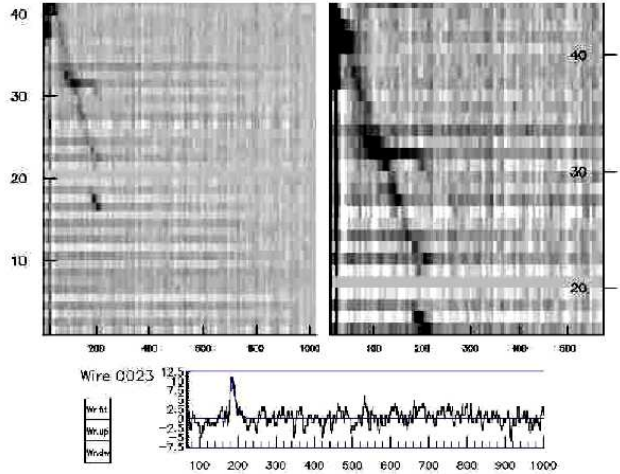}
\caption{A cosmic ray event in the Yale LArTPC. Left: Wire number versus time (sample number) in the collection plane. The wire pitch is 5~mm and the time is in units of 0.4~$\mu$s, corresponding to 0.2~mm for the applied field of 100~V/cm. Right: A zoomed-in view of the upper portion of the track.  Bottom: The waveform from wire $\#$23 in the collection plane.}
\label{fig:YaleTrack}
\end{figure}

Observing cosmic ray tracks was used as a practical demonstration of the filter's ability to purify liquid argon. A FNAL-built activated-copper filter (without vacuum insulation) was used by the group at Yale to purify liquid argon (without a molecular sieve filter on the fill line) for a TPC with 33~cm diameter and 16~cm drift length.  The purity proved adequate as cosmic ray tracks were readily detected. The TPC can be seen in Figure~\ref{fig:YaleTPC} and an example event is shown in Figure~\ref{fig:YaleTrack}. A loss of drifting charge due to impurity attachment of less than 40\% over a drift time of 0.5~ms (equivalent to an electron lifetime better than 1~ms) has been repeatedly measured in the TPC vessel, stable over a period of 24~hours without recirculation of the argon through a purification system. A detailed description of the test stand can be found in~\cite{Curioni:2007}

\begin{figure}
\centering
\includegraphics[width=1.4in,viewport=260 220 520 620]{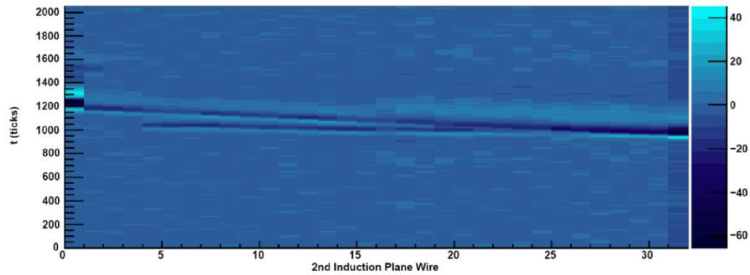}
\caption{A cosmic ray event in the FNAL LArTPC. The display shows time (sample number) versus wire number in an induction plane. The wire pitch is 4.7~mm and the time is in units of 198~ns, corresponding to 0.3~mm for the applied field of 500~V/cm.}
\label{fig:BoTrack}
\end{figure}

A LArTPC at FNAL has also recently detected cosmic ray tracks. Figure~\ref{fig:BoTrack} shows a through-going cosmic ray muon (with visible delta ray) on an induction plane in the test stand. The regenerable filter design described in this work was used for the detector's liquid argon purification system as well. 

\section{Conclusion}
We have constructed and regenerated filters capable of removing electronegative impurities from liquid argon. The methods described here purify liquid argon to the 10's of ppt level, corresponding to many millisecond electron drift lifetimes.  The simply constructed, regenerable, and non-proprietary filters provide an efficient and practical method to purify commercial-grade liquid argon for use in LArTPCs. 

\section{Acknowledgements}
The electronics and data-acquisition system for the TPC at FNAL were designed and built by D. Edmunds and P. Laurens of Michigan State University. The Yale group acknowledges essential help from: N. Canci and F. Arneodo of LNGS, in the initial work on LAr purification; S. Centro, S. Ventura, B. Baibussinov and the ICARUS group at INFN Padova, for the readout electronics, software for DAQ and event display;  L. Bartoszek of Bartoszek Engineering. This work is supported by the Department of Energy through FNAL and the Advanced Detector Research Program, and through the National Science Foundation.
\newpage
\bibliographystyle{elsarticle-num}
\bibliography{Purity.bib}
\end{document}